\documentclass[aps,prd,amsmath
,eqsecnum            
,preprintnumbers
,nofootinbib         
 ,showpacs          
 ,showkeys          
,twocolumn         
]{revtex4}
\usepackage[dvips]{graphicx}
\usepackage{epsfig}

\usepackage{amsmath}
\usepackage{amsfonts}
\usepackage{amssymb}
\usepackage{longtable}   
\usepackage{rotating}

\allowdisplaybreaks[4]     

\newcommand{\df}{\ {\overset {\rm def} =}\ }
\newcommand{\dr}[2]{\frac {{\rm d} {#1}} {{\rm d} {#2}}}
\newcommand{\pdr}[2]{\frac {\partial {#1}} {\partial {#2}}}
\newcommand{\dril}[2]{{{\rm d} {#1}} / {{\rm d} {#2}}}

\begin{document}

\title{Apparent horizons in the quasispherical Szekeres models}


\author{Andrzej Krasi\'nski}
\affiliation{N. Copernicus Astronomical Centre, Polish Academy of Sciences, \\
Bartycka 18, 00 716 Warszawa, Poland} \email{akr@camk.edu.pl}
\author{Krzysztof Bolejko}
\affiliation{Astrophysics Department, University of Oxford, Oxford OX1 3RH, UK}
\email{Krzysztof.Bolejko@astro.ox.ac.uk}

\date {}

\begin{abstract}
The notion of an apparent horizon (AH) in a collapsing object can be carried
over from the Lema\^{\i}tre -- Tolman (L--T) to the quasispherical Szekeres
models in three ways: 1. Literally by the definition -- the AH is the boundary
of the region, in which every bundle of null geodesics has negative expansion
scalar. 2. As the locus, at which null lines that are as nearly radial as
possible are turned toward decreasing areal radius $R$. These lines are in
general nongeodesic. The name ``absolute apparent horizon'' (AAH) is proposed
for this locus. 3. As the boundary of a region, where null \textit{geodesics}
are turned toward decreasing $R$. The name ``light collapse region'' (LCR) is
proposed for this region (which is 3-dimensional in every space of constant
$t$); its boundary coincides with the AAH. The AH and AAH coincide in the L--T
models. In the quasispherical Szekeres models, the AH is different from (but not
disjoint with) the AAH. Properties of the AAH and LCR are investigated, and the
relations between the AAH and the AH are illustrated with diagrams using an
explicit example of a Szekeres metric. It turns out that an observer who is
already within the AH is, for some time, not yet within the AAH. Nevertheless,
no light signal can be sent through the AH from the inside. The analogue of the
AAH for massive particles is also considered.
\end{abstract}

\maketitle

\section{Motivation}

This paper deals with the relationship between analogues of an apparent horizon
(AH) that exist in the quasispherical Szekeres models \cite{Szek1975a} --
\cite{SuBo2012} of the $\beta' \neq 0$ family\footnote{In the literature there
are several definitions of different kinds of horizons. Some of them require
asymptotic flatness, others noninteraction with the surroundings. These do not
apply here because of the dynamical character of the Szekeres spacetime. For
different types of horizons and discussion the reader is referred to
\cite{Hayw1994,Seno2012}.}. The AH was first defined by Hawking and Ellis (HE,
\cite{HaEl1973}, in what follows we quote from this source) as the outer
boundary of a connected component of an outer trapped region within a partial
Cauchy surface ${\cal S}(\tau)$. A trapped region is the collection of all
points $q \in {\cal S}(\tau)$ such that there exists an outer trapped surface
${\cal P} \subset {\cal S}(\tau)$ containing $q$. An outer trapped surface is a
2-surface in ${\cal S}(\tau)$ such that the family of outgoing null geodesics
orthogonal to it has nonpositive expansion scalar, as defined by Sachs
\cite{PlKr2006}. In our case, the partial Cauchy surfaces will be the
hypersurfaces of constant $t$. It is a simple and rather small step forward from
this definition to consider the collection of all apparent horizons in the sense
of HE, and retain the name AH for the 3-dimensional hypersurface thus formed. We
shall use the term AH in the latter sense. Szekeres \cite{Szek1975b} showed that
in his quasispherical model the AH in this broader sense is located at $R = 2M$
(in our notation). (He did not use the term ``AH''.) Hellaby and Krasi\'nski
\cite{HeKr2002} gave the name of AH to a different entity, for which the name
``absolute apparent horizon'' (AAH) is proposed here. The AAH is defined in
terms of \textit{nongeodesic} null lines that are, in a sense to be defined in
Sec. \ref{ahvsaah}, as nearly radial as possible. (Strictly radial curves do not
exist in general Szekeres models because of their lack of symmetry
\cite{NoDe2007}.) The AAH is the locus at which these nearly radial null curves
are turned toward decreasing areal radius $R$. In the Lema\^{\i}tre
\cite{Lema1933} -- Tolman \cite{Tolm1934} (L--T) model, which is the spherically
symmetric limit of the Szekeres models considered here, the AAH coincides with
the AH \cite{KrHe2004}, and the curves defining the AAH become radial null
geodesics.

One more analogue of AH results when we consider null geodesics and the region,
where they are turned toward decreasing $R$. For this region, the name ``light
collapse region'' (LCR) is proposed here. Unlike AAH and AH, which are
3-dimensional hypersurfaces in spacetime, the LCR is a 4-dimensional subset of
spacetime because the family of geodesics defining it is not uniquely
determined.

The existence of the AAH is proven for every collapsing quasispherical Szekeres
model in Sec. \ref{ahvsaah}. In Sec, \ref{LCRsec}, the LCR is defined and it is
shown that the 3-dimensional future boundary of LCR coincides with the AAH. In
Sec. \ref{AAHexpli} an explicit subcase of the quasispherical Szekeres model is
chosen for a detailed investigation. We illustrate the relation between the AH
and the AAH by diagrams showing their positions in space. It turns out that, for
some directions, an observer who is already within the AH is, for some time, not
yet within the AAH. The analogue of AAH for massive particles is also
considered. In Sec. \ref{truehor} the matching of the quasispherical Szekeres
solutions to the Schwarzschild solution is considered. It is shown that it is
the AH that matches to the Schwarzschild event horizon located at $r = 2m$ and
that the outgoing part of the AH is necessarily spacelike, so light rays cannot
traverse it outwards from the inside. Both these facts indicate that the AH
rather than the AAH is the true horizon.

The aim of this paper is to gain more insight into the geometry of the Szekeres
solutions.

\section{Introducing the $\beta' \neq 0$ quasispherical Szekeres
solutions}\label{intszek}

\setcounter{equation}{0}

In this section, basic facts about the $\beta' \neq 0$ quasispherical Szekeres
solutions are recalled for reference, and to define the notation. We will use
the parametrization introduced by Hellaby \cite{Hell1996}. The metric of these
solutions is
\begin{eqnarray}\label{2.1}
&& {\cal E} \df \frac S 2 \left[\left(\frac {x - P} S\right)^2 + \left(\frac
{y - Q} S\right)^2 + 1\right], \\
&& {\rm d} s^2 = {\rm d} t^2 - \frac {\left(R,_z - R {\cal E},_z/{\cal
E}\right)^2} {1 + 2E(z)} {\rm d} z^2 - \frac {R^2} {{\cal E}^2} \left({\rm d}
x^2 + {\rm d} y^2\right).\nonumber
\end{eqnarray}
where $E(z), P(z), Q(z), S(z)$ are arbitrary functions, and $R(t, z)$ obeys the
following equation; a consequence of Einstein's equations with dust source:
\begin{equation}\label{2.2}
{R,_t}^2 = 2E(z) + \frac {2M(z)} R + \frac 1 3 \Lambda R^2;
\end{equation}
$M(z)$ being one more arbitrary function, and $\Lambda$ being the cosmological
constant. The coordinates of (\ref{2.1}) are comoving, so the velocity field of
the dust is $u^{\mu} = {\delta^{\mu}}_0$ and $\dot{u}^{\mu} = 0$. In the
following we assume $\Lambda = 0$.

This solution has in general no symmetry, and reduces to the L--T solution when
$P$, $Q$ and $S$ are all constant.\footnote{The ``$\beta' \neq 0$'' refers to
the fact that ${\rm e}^{\beta} \df R / {\cal E}$ depends on $z$, so $\beta'
\equiv \beta,_z \neq 0$ (this notation follows Szekeres
\cite{Szek1975a,Szek1975b}). There exists another large family of Szekeres
solutions, in which $\beta' = 0$. They require separate treatment and will not
be considered here. See an extended presentation in Ref. \cite{PlKr2006}, also
for the associated quasiplane and quasihyperbolic Szekeres models.} The sign of
$E(z)$ determines the type of evolution; with $E(z_0) < 0$ the matter shell at
$z = z_0$ expands away from an initial singularity and then recollapses to a
final singularity, with $E(z_0) > 0$ the shell is ever-expanding or
ever-collapsing, depending on the initial conditions; $E(z_0) = 0$ is the
intermediate case, ever-expanding with asymptotically zero expansion velocity,
or its time-reverse. All three evolution types may exist in different regions of
the same spacetime. We consider here the recollapsing ($E < 0$) solution of
(\ref{2.2}) with $\Lambda = 0$,
\begin{eqnarray}\label{2.3}
R &=& - \frac M {2E} (1 - \cos \eta), \nonumber \\
\eta - \sin \eta &=& \frac {(- 2E)^{3/2}} M\ \left(t - t_B(z)\right),
\end{eqnarray}
where $t_B(z)$ is one more arbitrary function and $\eta(t, z)$ is a parameter.
The mass-density in energy units is
\begin{equation}\label{2.4}
\kappa \epsilon = \frac {2 \left(M,_z - 3 M {\cal E},_z / {\cal E}\right)} {R^2
\left(R,_z - R {\cal E},_z / {\cal E}\right)}, \qquad \kappa \df \frac {8 \pi G}
{c^4}.
\end{equation}
For $\epsilon > 0$, $\left(M,_z - 3 M {\cal E},_z / {\cal E}\right)$ and
$\left(R,_z - R {\cal E},_z / {\cal E}\right)$ must have the same sign. Note
that the sign may be flipped by the transformation $z \to - z$, so we may assume
that ${\cal L} \df R,_z - R {\cal E},_z / {\cal E} > 0$ at least somewhere. Let
us then consider whether ${\cal L}$ can change sign as a function of $z$. The
set where ${\cal L} = 0$ is either (1) a curvature singularity (a shell crossing
-- see a comment on it in the next section) or (2), if it coincides with the set
where $M,_z - 3 M {\cal E},_z / {\cal E} = 0$, an analogue of a neck, well-known
from the L--T geometry \cite{PlKr2006}. Case (1) is excluded by assumption -- we
choose the functions in the model so that shell crossings do not occur. In case
(2), the neck (if it exists) is a global feature of spacetime, and no ordinary
astronomical object in our neighbourhood is large enough to extend up to and
through it. Thus, we assume that we are on one side of the neck, where
\begin{equation}\label{2.5}
R,_z - R {\cal E},_z / {\cal E} > 0 \Longrightarrow M,_z - 3 M {\cal E},_z /
{\cal E} > 0.
\end{equation}
We also assume
\begin{equation}\label{2.6}
M,_z > 0
\end{equation}
because the region where $M,_z < 0$ occurs is an analogue of the region behind
the equator of a closed space in the positive-curvature Robertson--Walker
spacetimes. Again, no astronomical object is that large.

The Robertson--Walker limit follows when $z = r$, $R (t,z) = r S(t)$, $E = E_0
r^2$, where $E_0 =$ constant and $P = Q = 0$, $S = 1$. This definition includes
the definition of the limiting radial coordinate (the Szekeres model is
covariant with the transformations $z = f(z')$, where $f(z')$ is an arbitrary
function).

The quasispherical model may be imagined as such a generalisation of the L--T
model in which the spheres of constant mass were made nonconcentric. The
functions $P(z)$, $Q(z)$ and $S(z)$ determine how the center of a sphere changes
its position in a space $t =$ constant when the radius of the sphere is
increased or decreased \cite{HeKr2002}.

Within each single $\{t =$ constant, $z =$ constant$\}$ surface, which is a
sphere, the $(x, y)$ coordinates of (\ref{2.1}) can be transformed to the
spherical $(\vartheta, \varphi)$ coordinates by
\begin{equation}\label{2.7}
(x - P, y - Q) / S = \cot (\vartheta / 2) (\cos \varphi, \sin \varphi).
\end{equation}
This transformation is called a {\it stereographic projection}. For its
geometric interpretation see Refs. \cite{HeKr2002} and \cite{PlKr2006}. Using
this transformation the factor $ {\cal E},_z / {\cal E}$ becomes
\begin{equation}\label{2.8}
{\cal E},_z / {\cal E} = -\left[S,_z \cos \vartheta + \sin \vartheta \left( P,_z
\cos \varphi + Q,_z \sin \varphi \right)\right]/S.
\end{equation}

\section{Properties of the quasispherical Szekeres solutions}
\label{szeksolprop}

\setcounter{equation}{0}

Definitions of the Szekeres solutions based on invariant properties can be found
using Ref. \cite{PlKr2006}.

Rotation and acceleration of the dust source are zero, the expansion is nonzero,
the shear tensor is
\begin{eqnarray}\label{3.1}
{\sigma^{\alpha}}_{\beta} &=& \frac 1 3 \Sigma \  {\rm diag\ } (0, 2, -1, -1),
\qquad {\rm where} \qquad \nonumber \\
\Sigma &=& \frac {R,_{tz} - R,_t R,_z / R} {R,_z - R {\cal E},_z / {\cal E}}.
\end{eqnarray}

The instant $t = t_B(z)$ in (\ref{2.3}) is the Big Bang singularity
corresponding to $R = 0$. When $t_{B,z} \neq 0$ (that is, in general) the
instant of singularity is position-dependent.

Another singularity may occur where $R,_z - R {\cal E},_z / {\cal E} = 0$ (if
this equation has solutions for $(x ,y)$). This is a shell crossing, but it is
qualitatively different from that in the L--T model. As can be seen from
(\ref{2.1}), the equation $R,_z - R {\cal E},_z / {\cal E} = 0$ can define at
most a subset of an $\{x, y\}$ sphere. When a shell crossing exists, its
intersection with a $t =$ constant space will be a circle, or, in exceptional
cases, a single point (in L--T it is a whole sphere). For more on shell
crossings in all the Szekeres solutions see Refs. \cite{HeKr2002} and
\cite{HeKr2008}. They can be avoided if the functions and their derivatives obey
a set of inequalities \cite{HeKr2002,SuBo2012}.

Equation (\ref{2.2}) is formally identical with the Friedmann equation, but,
with $E$ and $M$ depending on $z$, each surface $z$ = constant evolves
independently of the others.

A quasispherical Szekeres region can be matched to the Schwarzschild solution
across a $z =$ constant hypersurface \cite{Bonn1976a}.

The mass-density distribution given by (\ref{2.4}) can be decomposed into the
spherically symmetric monopole
\begin{equation}\label{3.2}
\kappa \epsilon_S = \frac {2 (M/\chi)^3} {(R/\chi)^2 (R/\chi),_z},
\end{equation}
where
\begin{equation}\label{3.3}
\chi(z) \df \frac {P^2 + Q^2 + S^2 + 1} S,
\end{equation}
and the dipole
\begin{equation}\label{3.4}
\kappa \Delta \epsilon = \frac {6MR,_z - 2M,_z R} {R^2 \left(R,_z \chi - R
\chi,_z\right)} \times \frac {\chi,_z - \chi {\cal E},_z/{\cal E}} {R,_z - R
{\cal E},_z / {\cal E}}.
\end{equation}
The dipole is uniquely defined by the requirement that the surface where $\Delta
\epsilon = 0$ (sure to exist, as follows from calculations -- see Refs.
\cite{BKHC2010} and \cite{PlKr2006}) passes through the center of symmetry of
the monopole.\footnote{Equation (\ref{3.4}) corrects a typo in eqs. (2.194) and
(2.196) of Ref. \cite{BKHC2010} and in eq. (19.165) of Ref. \cite{PlKr2006}: one
of the two appearances of $\Phi^2$ in each equation should not be there.}

\section{Apparent vs absolute apparent horizons in the quasispherical Szekeres
models} \label{ahvsaah}

\setcounter{equation}{0}

The results of this section were partly reported in Ref. \cite{BKHC2010}; the
basic equations were introduced in Ref. \cite{HeKr2002}.

An AH is the boundary of the region of trapped surfaces. A trapped surface $S$
is one on which the families of outgoing null \textit{geodesics} on both sides
of $S$ converge (i.e. have a negative expansion scalar). Thus, if $k^{\mu}$ is
any field of vectors tangent to null geodesics that intersect $S$, then
\begin{equation}\label{4.1}
{k^{\mu}};_{\mu} < 0 \qquad {\rm on\ } S.
\end{equation}
Consequently, on an AH:
\begin{equation}\label{4.2}
{k^{\mu}};_{\mu} = 0.
\end{equation}
Proceeding from this definition, Szekeres \cite{Szek1975b} found that in a
quasispherical model the AH is given by the same equation as in an L--T model:
\begin{equation}\label{4.3}
R = 2M.
\end{equation}
In an L--T model, (\ref{4.2}) is equivalent to another definition: on an AH in
collapsing matter, $R(z)$ calculated along an outgoing radial null geodesic
changes from increasing to decreasing \cite{PlKr2006}.

Hellaby and Krasi\'nski \cite{HeKr2002} considered the analogue of an AH in a
quasispherical Szekeres model, using this second definition, but for {\it
nongeodesic} null fields defined below. We propose to name this the ``absolute
apparent horizon'' (AAH) -- because even a maximally accelerated ray cannot get
out of it.

The reasoning was as follows. A general null direction $k^{\alpha} = \dril
{x^{\alpha}} t$ in the metric (\ref{2.1}) obeys\footnote{Along a null curve
parametrized by an affine parameter $s$, the time coordinate $t$ must obey ${\rm
d}t/{\rm d}s > 0$ or ${\rm d}t/{\rm d}s < 0$ at all points (the curve would be
spacelike at every point where ${\rm d}t/{\rm d}s = 0$). This shows that $t$ can
be used as a parameter on null geodesics (but in general it is not affine).}
\begin{equation}\label{4.4}
\frac{\left(R,_z - R {\cal E},_z / {\cal E}\right)^2} {1 + 2E} \left(\dr z
t\right)^2 = 1 - \frac {R^2} {{\cal E}^2} \left[\left(\dr x t\right)^2 +
\left(\dr y t\right)^2\right].
 \end{equation}
Thus, on a null curve with $\dril x t = 0 = \dril y t$ (which, in general, will
not be a geodesic \cite{BKHC2010,KrBo2011,NoDe2007}), $\left|\dril z t\right|$
is maximal. Equation (\ref{4.4}) implies, along this path:
\begin{equation}\label{4.5}
\left. \dr t z\right|_n = \frac j {\sqrt{1 + 2E}} \left(R,_z - \frac {R {\cal
E},_z} {\cal E}\right), \qquad j = \pm 1,
\end{equation}
where $j = +1$ for outgoing rays, and $j = -1$ for ingoing rays. Intuition
suggests that along a curve (\ref{4.5}) the light signal should escape farther
from the ``origin'' $R = 0$ than along any other path. An example in Sec.
\ref{AAHexpli} will show that this is only partly true. Along some directions,
the rays given by (\ref{4.5}) can indeed proceed toward larger values of $R$
even within the AH. But along the other directions the reverse happens: the rays
(\ref{4.5}) are redirected to decreasing values of $R$ even outside the AH. The
reason for this behaviour is the fact that the two definitions of AH that are
equivalent in the L--T limit are \textit{inequivalent} in a Szekeres model: the
locus where all bundles of light rays begin to converge is different from the
locus where light rays are forced to collapse toward decreasing $R$, see Sec.
\ref{LCRsec}.

Let the solution of (\ref{4.5}) be
 \begin{equation}\label{4.6}
   t = t_n(z).
 \end{equation}
The value of $R$ along this ray, $R_n(z) \df$ $R(t_n(z), z)$, is a monotonic
function of $z$ in some neighbourhood of the emission point. The AAH is where
$R_n(z)$ changes from increasing to decreasing or vice versa:
\begin{eqnarray}\label{4.7}
&& 0 = \dr {R_n} z \equiv \pdr R t \dr {t_n} z + \pdr R z \nonumber \\
&& = \ell j \frac {\sqrt{2 M / R + 2E}} {\sqrt{1 + 2E}} \left(R,_z - \frac {R
{\cal E},_z} {\cal E}\right) + R,_z,
 \end{eqnarray}
from (\ref{2.2}) and (\ref{4.5}), where $\ell = +1$ for an expanding model and
$\ell = -1$ for a collapsing model. We consider
 \\
 \\
 \\
an AAH that is created in the collapse phase ($\ell = -1$), so it is defined by
the outgoing rays ($j = +1$). Then, (\ref{4.7}) becomes
\begin{eqnarray}\label{4.8}
&& R,_z \left(\frac {\sqrt{1 + 2E}} {\sqrt{2M/R + 2E}} - 1\right) + R \frac
{{\cal E},_z} {\cal E} = 0. \\
&& \nonumber \\
&& \nonumber
\end{eqnarray}

\vspace{-8mm}

In Ref. \cite{HeKr2002} it was found that in a constant-$t$ space the AAH ``is a
kind of oval with half inside $R = 2M$ and half outside''.\footnote{This can be
easily seen from (\ref{4.8}). Suppose, for definiteness, that $R,_z
> 0$. Recall that ${\cal E} > 0$ (evident from (\ref{2.1})). Then $({\cal E},_z
> 0) \Longrightarrow (R < 2M)$, $({\cal E},_z = 0) \Longrightarrow (R = 2M)$
and $({\cal E},_z < 0) \Longrightarrow (R > 2M)$. For the proof that ${\cal
E},_z$ changes sign on an $(x, y)$ sphere see Ref. \cite{HeKr2002}; ${\cal E},_z
= 0$ is a large circle on that sphere.} In Sec. \ref{AAHexpli} we will
investigate the relation of the AAH to the AH in a simple example of a
recollapsing Szekeres model, and this will provide an illustration to the quoted
statement.

We use the following expression for $R,_z$ (to be calculated from (\ref{2.3});
see eq. (18.107) in Ref. \cite{PlKr2006}):
\begin{eqnarray}\label{4.9}
&& \frac {R,_z} R = \left(\frac {M,_z} M - \frac {E,_z} E\right) + \left(\frac
3 2 \frac {E,_z} E - \frac {M,_z} M\right)  \nonumber \\
&& \nonumber \\
&&\times \frac {\sin \eta (\eta - \sin \eta)} {(1 - \cos \eta)^2} - \frac {(-
2E)^{3/2}} M t_{B,z} \frac {\sin \eta} {(1 - \cos \eta)^2}.\ \ \ \
\end{eqnarray}
We note from (\ref{2.2}) and (\ref{2.3}) that with $\pi \leq \eta \leq 2\pi$,
where $R,_t < 0$, we have
\begin{equation}\label{4.10}
\sqrt{2M/R + 2E} = - \sqrt{- 2E} \frac {\sin \eta} {1 - \cos \eta}.
\end{equation}
We substitute (\ref{4.9}) and (\ref{4.10}) in (\ref{4.8}), then multiply the
result by $(1 - \cos \eta)^2$ to avoid the infinite values at $\eta \to 0$ and
$\eta \to 2 \pi$, and obtain:
 \\
 \begin{widetext}
\begin{eqnarray}\label{4.11}
\Psi (\eta) &\df& \left[\left(\frac {M,_z} M - \frac {E,_z} E\right) (1 - \cos
\eta)^{3/2} + \left(\frac 3 2 \frac {E,_z} E - \frac {M,_z} M\right) \frac {\sin
\eta (\eta - \sin \eta)} {\sqrt{1 - \cos \eta}} - \frac {(- 2E)^{3/2}} M t_{B,z}
\frac {\sin \eta} {\sqrt{1 - \cos \eta}}\right] \nonumber \\
&\times& \left[\sqrt{1 + 2E} \sqrt{1 - \cos \eta} + \frac {\sqrt{- 2E} \sin
\eta} {\sqrt{1 - \cos \eta}}\right] - \sqrt{- 2E} \sin \eta (1 - \cos \eta)
\frac {{\cal E},_z} {\cal E} = 0.
\end{eqnarray}
 \end{widetext}
This determines $\eta(M, x, y)_{\rm AAH}$. Then, from (\ref{2.3}):
\begin{equation}\label{4.12}
t(M, x, y)_{\rm AAH} = \frac M {(- 2E)^{3/2}}\ (\eta - \sin \eta)_{\rm AAH} +
t_B.
\end{equation}

We assume that shell crossings are absent. Among the conditions for no shell
crossings, found in Ref. \cite{HeKr2002}, the following are useful here:
\begin{equation}\label{4.13}
2 \pi \left(\frac 3 2 \frac {E,_z} E - \frac {M,_z} M\right) - \frac {(-
2E)^{3/2}} M t_{B,z} < 0
\end{equation}
(see eq. (126) in Ref. \cite{HeKr2002}), and
\begin{equation}\label{4.14}
M,_z / M - E,_z / E > 0,
\end{equation}
which follows from the fact that $R,_z/R > 0$ must hold for all $(\eta, z)$, via
(\ref{4.9}) taken at $\eta = \pi$ \cite{HeKr2002}. We observe that
\begin{equation}\label{4.15}
- \lim_{\eta \to 2\pi} \frac {\sin \eta} {\sqrt{1 - \cos \eta}} = \sqrt{2} =
\lim_{\eta \to 0} \frac {\sin \eta} {\sqrt{1 - \cos \eta}}.
\end{equation}
Now we verify using (\ref{4.11}) that
 \\
\begin{equation}\label{4.16}
\lim_{\eta \to \pi} \Psi(\eta) = 4 \sqrt{1 + 2E}\left(\frac {M,_z} M - \frac
{E,_z} E\right) > 0,
\end{equation}
being positive in consequence of (\ref{4.14}); and
\begin{eqnarray}\label{4.17}
&&\lim_{\eta \to 2\pi} \Psi(\eta) \\
&& = 2 \sqrt{- 2E} \left[2 \pi \left(\frac 3 2 \frac {E,_z} E - \frac {M,_z}
M\right) - \frac {(- 2E)^{3/2}} M t_{B,z}\right] < 0, \nonumber
\end{eqnarray}
being negative in consequence of (\ref{4.13}).

Thus $\Psi(\pi) > 0$ and $\Psi(2 \pi) < 0$, so there exists an $\eta_0 \in (\pi,
2 \pi)$ at which $\Psi(\eta_0) = 0$, and it is unique (see Appendix
\ref{unique}). In passing, we have proved that each particle in a recollapsing
quasispherical Szekeres model must cross the AAH before it hits the Big Crunch
at $\eta = 2 \pi$.

\section{The light collapse region (LCR) and its future boundary}\label{LCRsec}

\setcounter{equation}{0}

Consider a bundle of geodesic light rays flashed simultaneously from a common
origin. Let $v$ be the affine parameter along these rays, $\theta$ be the
expansion scalar of the bundle, $k^{\mu}$ be the tangent vector field to the
rays and $\delta S$ be the surface area of the propagating front of the bundle.
Then the following holds (\cite{PlKr2006}, eq. (16.131)):
\begin{equation}\label{5.1}
{k^{\mu}};_{\mu} \equiv 2 \theta = \dr {} v \ln (\delta S).
\end{equation}
Consequently, by (\ref{4.2}), on an AH the $\delta S$ stops increasing along the
geodesics in the bundle defining the AH and begins to decrease. In an L--T
model, a light front (LF) flashed from the origin $R = 0$ remains spherically
symmetric at all times, and its surface area is proportional to $R^2$ calculated
at the LF. Therefore, in a collapsing L--T model, the AH is at the same time the
locus at which $\left.R\right|_{\rm LF}$ reaches its maximum. \textit{This
coincidence between $\theta = 0$ and the maximum of $\left.R\right|_{\rm LF}$
does not hold in a Szekeres model}, as is demonstrated below. The LF is not
spherically symmetric, different points on it have different $R$ values at a
constant $t$, so the area of the front is no longer proportional to $R^2$. Thus,
there may be locations where $\theta < 0$, but $R$ is still increasing along the
rays, and locations where $\theta > 0$ while $R$ is decreasing. This remark
should help in understanding the relation between AH and AAH discussed in Sec.
\ref{AAHexpli}.

This noncoincidence allows us to define one more entity related to AH, which we
propose to name the ``light collapse region'' (LCR). This is the region where
$R$ has extrema along null geodesics, and so, during collapse, the rays are
turned toward the Big Crunch.

Consider (\ref{4.4}) along a null geodesic, and suppose we know the solution of
the geodesic equations. Then (\ref{4.4}) together with the geodesic equations
defines the function
\begin{equation}\label{5.2}
t = t_{ng}(z),
\end{equation}
where ``ng'' stands for ``along a null geodesic''. For an outward-directed null
geodesic we then have
\begin{equation}\label{5.3}
\left. \dr t z\right|_{ng} = \left.\frac {R,_z - R {\cal E},_z / {\cal E}}
{\sqrt{1 + 2E}\ U}\right|_{ng},
\end{equation}
where
\begin{equation}\label{5.4}
U \df \left.\sqrt{1 - \frac {R^2} {{\cal E}^2} \left[\left(\dr x t\right)^2 +
\left(\dr y t\right)^2\right]}\right|_{ng}.
\end{equation}
Proceeding exactly as from (\ref{4.6}) to (\ref{4.8}), and assuming we consider
outward-directed null geodesics in the collapse phase of the model we arrive at
the following analogue of (\ref{4.8})
\begin{eqnarray}\label{5.5}
&& \left.\dr R z\right|_{ng} = \frac {\sqrt{2M/R + 2E}} {\sqrt{1 + 2E}\ U}
\nonumber \\
&&\ \ \ \  \times \left.\left[R,_z \left(\frac {\sqrt{1 + 2E}\ U} {\sqrt{2M/R +
2E}} - 1\right) + R \frac {{\cal E},_z} {\cal E}\right]\right|_{ng}.\ \ \ \ \ \
\
\end{eqnarray}
We define the LCR as the region where
\begin{equation}\label{5.6}
\left.\dr R z\right|_{ng} = 0.
\end{equation}
As stated before, the LCR is a 4-dimensional subset of spacetime and a
3-dimensional subset of a space of constant time. This is because the geodesics
that define the LCR are not uniquely determined: the $U$ in (\ref{5.4}) depends
on the direction of the geodesic considered and takes a range of values at a
given $(t, z)$.

Consider the intersection of LCR with the AAH, i.e. a locus where (\ref{4.8})
and (\ref{5.6}) hold simultaneously. Within this set we have $U = 1
\Longrightarrow \dril x t = \dril y t = 0$. Thus, ${\rm LCR} \bigcap {\rm AAH}$
is a set in which both the nongeodesic null curves referred to in (\ref{4.5})
and the null geodesics referred to in (\ref{5.3}) -- (\ref{5.4}) begin to
proceed toward decreasing $R$, and in addition the null geodesics happen to have
$\dril x t = \dril y t = 0$ there, i.e. to be tangent to the curves defining the
AAH. The definition of this set is identical to the definition of AAH, eq.
(\ref{4.8}). This shows that the AAH is a boundary of the LCR. Actually, it is
the future boundary, as we show below. Consider the collection of null geodesics
that cross the AAH as defined by (\ref{4.8}). For them, calculate (\ref{5.5}) at
the points that obey (\ref{4.8}). It is convenient to rewrite (\ref{4.8}) and
(\ref{5.5}) as follows:
\begin{eqnarray}
&& R,_z - \left(R,_z - R {\cal E},_z / {\cal E}\right) \frac {\sqrt{2M/R + 2E}}
{\sqrt{1 + 2E}} = 0, \label{5.7} \\
&& \left.\dr R z\right|_{ng} = R,_z - \left(R,_z - R {\cal E},_z / {\cal
E}\right) \frac {\sqrt{2M/R + 2E}} {\sqrt{1 + 2E}\ U}. \ \ \ \ \ \ \label{5.8}
\end{eqnarray}
Using (\ref{5.7}) in (\ref{5.8}) we obtain
\begin{eqnarray}\label{5.9}
&& \beta \df \left.\dr R z\right|_{ng(AAH)} \nonumber \\
&& = \left(R,_z - R {\cal E},_z / {\cal E}\right) \frac {\sqrt{2M/R + 2E}}
{\sqrt{1 + 2E}} \left(1 - \frac 1 U\right).\ \ \ \ \
\end{eqnarray}
Using (\ref{2.5}), since $U \leq 1$ by construction, we see that $\beta \leq 0$.
Those geodesics, for which $\beta = 0$ ($U = 1$) cross the AAH with $\dril x t =
\dril y t = 0$ and are just being turned toward decreasing $R$. Those for which
$\beta < 0$, while crossing the AAH are already proceeding toward decreasing
$R$. This shows that the AAH lies at the future boundary of the LCR.

\section{Explicit examples of AAH in Szekeres models}\label{AAHexpli}

\subsection{Null rays}\label{AAHexpliA}

\setcounter{equation}{0}

As an illustration, we take the recollapsing Szekeres model defined by the same
equations that were used in Ref. \cite{KrHe2004} to discuss the formation of
galactic-size black holes in the L--T model:\footnote{The values of the
parameters $a, b, T_0$ and $t_{B0}$ used here will be different from those in
Ref. \cite{KrHe2004}. This model is meant to be an illustration to various
geometrical possibilities; it is not supposed to describe any real object in the
Universe.}
\begin{eqnarray}
t_B(M) &=& - b M^2 + t_{B0}, \label{6.1} \\
t_C(M) &=& a M^3 + T_0 + t_{B0}, \label{6.2}
 \end{eqnarray}
where $t_B(M)$ is the bang time, $t_C(M)$ is the crunch time, $a$, $b$, $t_{B0}$
and $T_0$ are arbitrary constants; $t_{B0}$ is the time-coordinate of the
central point of the Big Bang and $T_0$ is the time between the Big Bang and Big
Crunch measured along the central line $M = 0$. Then, from (\ref{2.3}), since
$\eta = 2 \pi$ at $t = t_C$:
 \begin{equation}\label{6.3}
2E(M) = - \frac {(2 \pi M)^{2/3}} {\left(aM^3 + bM^2 + T_0\right)^{2/3}}.
 \end{equation}

As shown in Ref. \cite{HeKr2002}, eq. (185), the extreme values of ${\cal E},_z
/ {\cal E}$ are
 \begin{equation}\label{6.4}
D_e \df \left. \frac {{\cal E},_z} {\cal E} \right|_{\rm extreme} = \pm
\frac{\sqrt{{S,_z}^2 + {P,_z}^2 + {Q,_z}^2}} S.
 \end{equation}

However, in choosing $(P, Q, S)$ precaution must be taken not to make $D_e$ too
large. If it is too large, then either the numerator or the denominator of
(\ref{2.4}) becomes negative in a region of space, thus rendering the mass
density negative there (and infinite where the denominator is zero). Physically,
this means that the dipole component of (\ref{3.4}) dominates over the monopole
in that part of the space.

\begin{figure}[h]
\begin{center}
 \includegraphics[scale = 0.7]{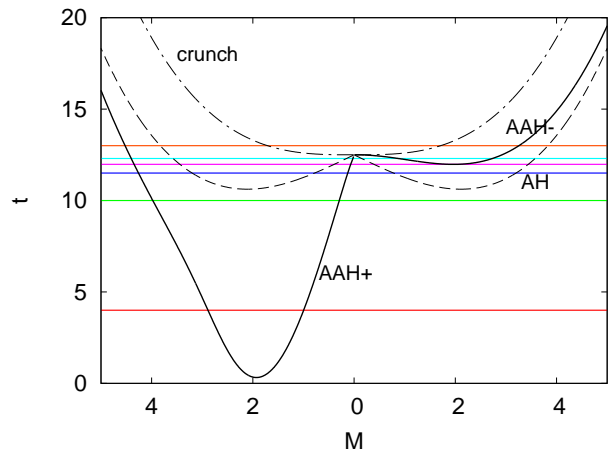}
\caption{A comparison of the future absolute apparent horizon (AAH) with the
ordinary future apparent horizon (AH) in the model defined by (\ref{6.1}) and
(\ref{6.2}). Curve AAH+ is the AAH along the direction where the contribution
from ${\cal E},_z / {\cal E}$ is maximal and curve AAH$-$ is the AAH along the
direction where this contribution is minimal. The dashed-dotted line represents
the big crunch singularity. Horizontal solid lines show the instants for which
the next figures are drawn, these are: $t = 4.0$, $10.0$, $11.5$, $11.9772$,
$12.3$ and $13.0$. Note that we use $M$ as the radial coordinate. }
\label{fig1}
\end{center}
\end{figure}

We thus first choose such a value of $D_e$ that will make the difference between
the graphs of AH and of AAH visible at the scale of a figure, and then we choose
such $P$, $Q$ and $S$ that will imply the chosen value of $D_e$. To maximize
$D_e$, at least one of the derivatives $P,_z$, $Q,_z$, $S,_z$ has to be large.
Experiments showed that the following functions will yield the desired result:
\begin{eqnarray}\label{6.5}
&&a = 0.1, \qquad b = 5000, \qquad T_0 = 12.5, \qquad t_{B0} = 0,\nonumber \\
&& S = M^{0.29} , \qquad P = 0.5 M^{0.29} , \qquad Q = 0,
\end{eqnarray}
and $z' = M(z)$ was chosen as the new $z$-coordinate. The resulting AAH is shown
in Fig. \ref{fig1}. The figure shows $t(M)$ on the AAH for two points in the
$(x, y)$ surface: the one where the $D_e$ given in (\ref{6.4}) is maximal
(positive) and where it is minimal (negative). These two curves are compared
with the ordinary AH and with the crunch time function $t_C(M)$. (This figure is
a modification of Fig. 1 in Ref. \cite{KrHe2004}.) See Appendix \ref{AAHat0} for
the proof that all four curves indeed have a common origin at $M = 0$.

Figure \ref{fig1} shows that the contribution from ${\cal E},_z / {\cal E}$ can
either increase or decrease the region where the accelerated rays are forced
toward the Big Crunch, depending on the direction. In the direction where this
contribution is maximal (i.e. ${\cal E},_z / {\cal E} > 0$ -- curve AAH$-$), the
AAH appears later than the ordinary AH, and the term ${\cal E},_z / {\cal E}$
causes that the accelerating ray can still proceed toward increasing $R$ in a
region where a geodesic bundle already converges. In the direction where $D_e$
is minimal (i.e. ${\cal E},_z / {\cal E} < 0$ -- curve AAH+), the AAH appears
earlier than the AH, and the term ${\cal E},_z / {\cal E}$ causes that the
accelerating ray is turned inward where a geodesic bundle is still diverging.

Note what this means physically. A nongeodesic light ray is one that is guided
by mirrors or optical fibers. When the AAH has a smaller radius than the AH, the
observer who has already fallen into the AH still has a chance to send a
message, using nongeodesic rays, to observers occupying loci with larger $R$.
However, the nongeodesic ray has no chance to escape from inside the AH and will
be turned toward the Big Crunch as well, only somewhat later than the geodesic
one -- see Sec. \ref{truehor}. Even this is possible only in some of the
directions; in other directions the AAH is outside the AH and no ray within the
AH, geodesic or not, can proceed toward larger $R$. (See later in this paper --
Figs. \ref{fig5} and \ref{fig6} illustrate this point more clearly.)

\begin{figure*}
  \includegraphics[scale = 1.0]{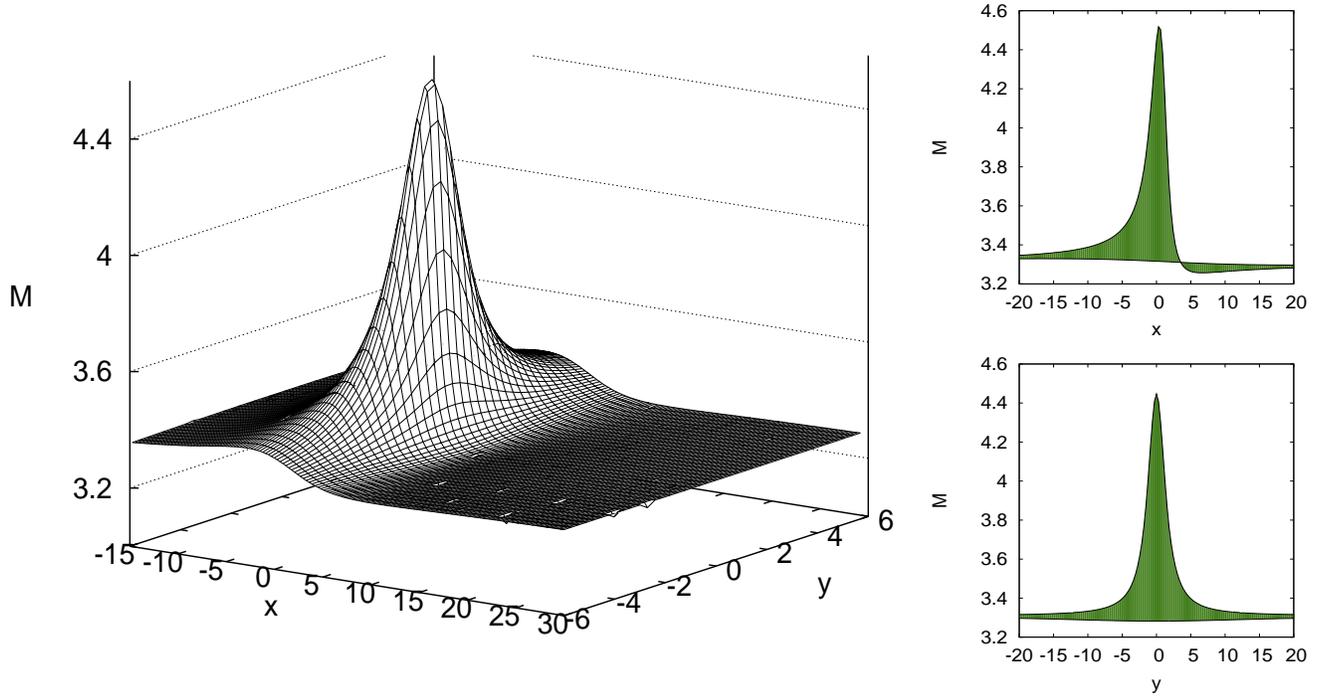}
\caption{Graph of $M(x, y)$ on the AAH in the space $t = 13.0$ (compare Fig.
\ref{fig1}). More explanation in the text.} \label{fig2}
\end{figure*}

\begin{figure*}
  \includegraphics[scale = 0.8]{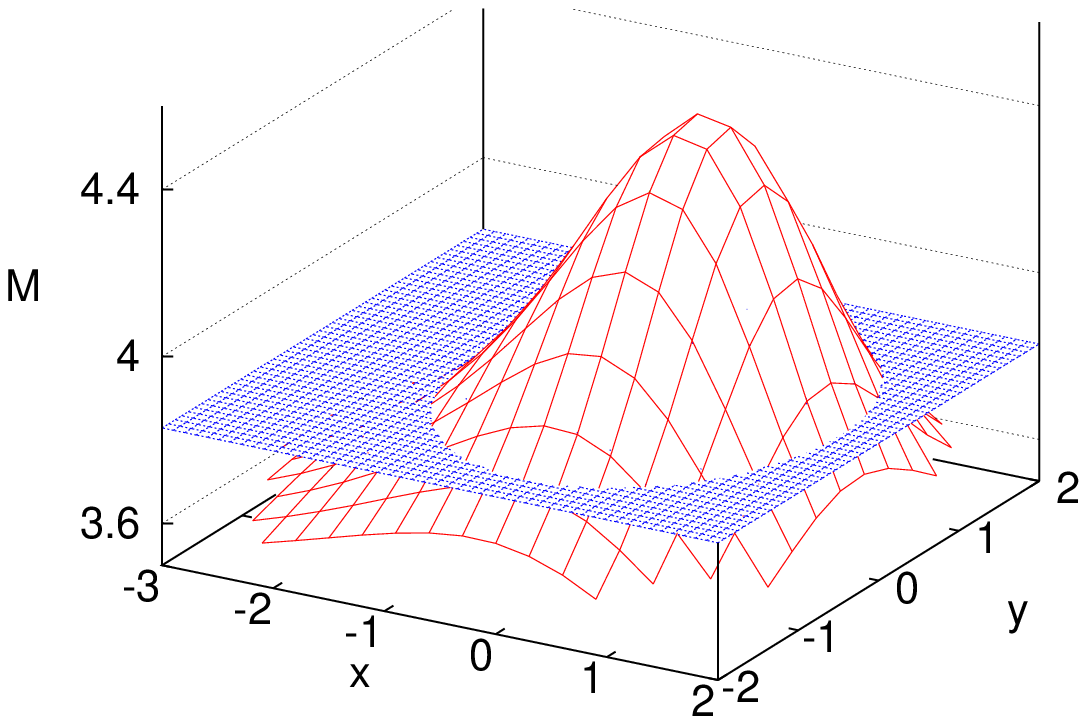}
  \includegraphics[scale = 0.55]{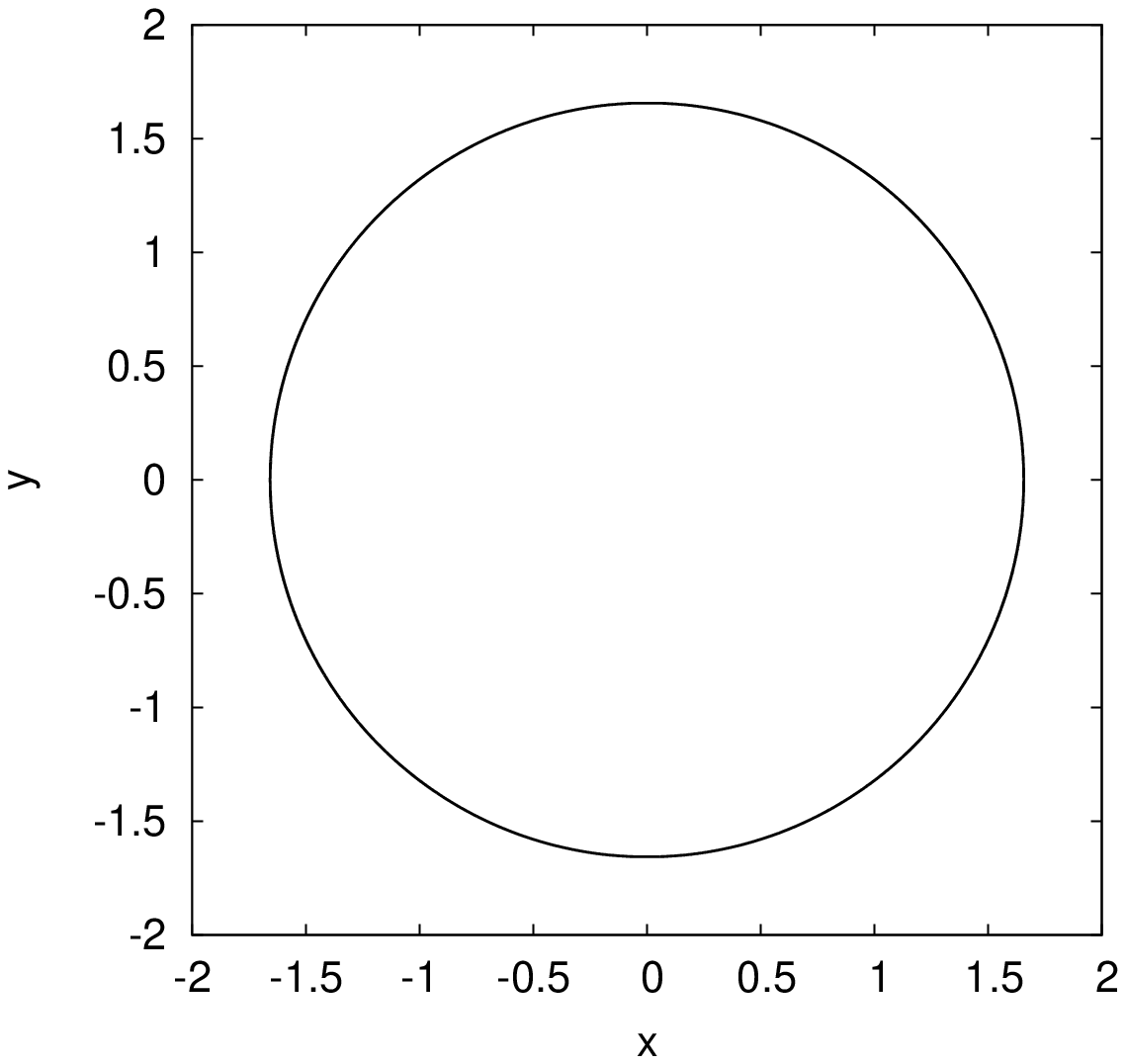}
\caption{{\em Left}: Intersection of the AAH with the ordinary AH (it lies in
the plane $M = 3.82860$) at $t = 13.0$.  {\em Right}: The line of intersection
of the AAH with the ordinary AH.} \label{fig3}
\end{figure*}

Figure \ref{fig2} shows a 3d graph of $M$ on the AAH as a function of $x$ and
$y$, at the time instant $t = 13.0$ (compare Fig. \ref{fig1}). All values of $x$
and $y$ are admissible, and at every pair $(x, y)$ there will be an $M$ obeying
(\ref{4.11}). The graph shows at which points in the $(x, y)$ plane the AAH has
the largest radius (as measured by $M$), and where it has the smallest radius.
The maximum of $M$ is at the intersection of curve AAH+ in Fig. \ref{fig1} with
the line $t = 13.0$; the minimum of $M$ is at the intersection of curve AAH$-$
with the same line. Comparison with Fig. \ref{fig1} shows that the values of $M$
are indeed all in the expected range. Figure \ref{fig3} shows the intersection
of the AAH with the ordinary AH, which, at $t = 13.0$ is at the mass
\begin{equation}\label{6.6}
M \df M_{\rm AH} = 3.82860
\end{equation}

\begin{figure*}
 \includegraphics[scale = 1.2]{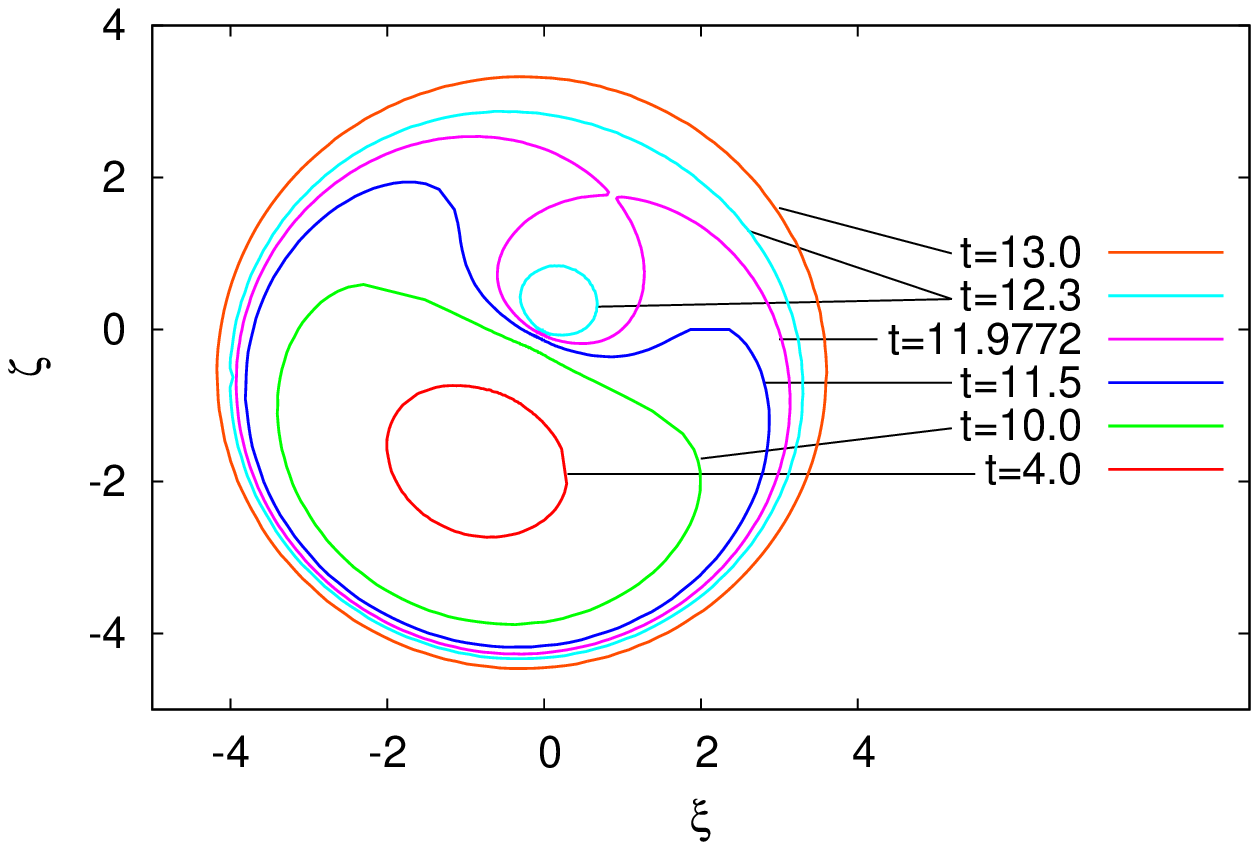}
\caption{The AAH at 6 different time instants (shown in Fig. \ref{fig1} as
horizontal lines). The curves shown are located on the plane $(\xi, \psi = 0,
\zeta)$. This figure can be imagined as a view of Fig. \ref{fig1} by an observer
sitting high on the $t$-axis and looking down; with one spatial dimension added.
The value of $M_{AAH}$ at a point $(\xi_0, \zeta_0)$ is the distance between
$(\xi_0, \zeta_0)$ and the origin $R = 0$, which is inside the smallest contour
in the upper right area at $(\xi, \zeta) = (0, 0)$. } \label{fig4}
\end{figure*}

The coordinates $(x,y)$ are not very intuitive. To better visualize the AAH let
us first use the stereographic projection (\ref{2.7}) to transform $(x,y)$ to
the $(\vartheta, \varphi)$ coordinates, and then map the AAH into an abstract
Euclidean space with the coordinates $(\xi, \psi, \zeta)$. The second
transformation has the following form
\begin{eqnarray}\label{6.7}
&& \xi = M_{AAH} (\vartheta, \varphi) \sin \vartheta \cos \varphi, \nonumber \\
&& \psi = M_{AAH} (\vartheta, \varphi) \sin \vartheta \sin \varphi, \nonumber \\
&& \zeta = M_{AAH} (\vartheta, \varphi) \cos \vartheta.
\end{eqnarray}
We now use these coordinates to present the evolution of the AAH. As seen from
(\ref{2.8}), when $Q_{,z} =0$ the extreme values of ${\cal E}_{,z}/{\cal E}$
with respect to $\varphi$ are when $\varphi = 0$ and $\varphi = \pi$, which, as
follows from (\ref{6.7}), implies  $\psi = 0$. Therefore, Fig. \ref{fig4}
presents the intersection of the AAH with the plane $(\xi, \psi = 0, \zeta)$ at
6 different time instants.

Note how the lack of spherical symmetry influences the situation. The
``origin'', where $R = 0$ ($\xi = 0$, $\zeta = 0$), is inside the smallest
contour in Fig. \ref{fig4}. The AAH+ first appears off the origin (inside the
closed curve in the lower left part of the figure). As seen, at this instant,
most rays will miss it. Then it increases in diameter and encroaches on the
origin. At the instant corresponding to the lowest point of AAH$-$ in Fig.
\ref{fig1} (see the purple curve in Fig. \ref{fig4}, which presents the AAH just
moments before this instant, $t = 11.9772$), the cross-section is still
connected, but consists of two tangent rings, one inside the other. The point of
tangency lies at the minimum of AAH$-$. From that moment on, the cross-section
splits into two disjoint contours, the smaller of which becomes progressively
smaller with increasing $t$, and shrinks to a point at the instant corresponding
to the minimum of the Big Crunch (inside the smallest ring in Fig. \ref{fig4}).

Figure \ref{fig4} does not show the cross-sections of the ordinary AH because
they would obscure the image. The AH first appears shortly after $t = 10.0$ and,
at the moment of first appearance, would show in Fig. \ref{fig4} as a single
circle with the center at $(\xi, \zeta) = (0, 0)$ and radius slightly larger
than $M = 2$. At later instants, the cross-section of the AH splits into two
circles, with the centers at the same point. The smaller circle has its radius
decreasing as $t$ increases, and shrinks to a point at $t = T_0 + t_{B0} = 12.5$
(the smaller contour of the AAH shrinks to a point at the same instant). The
larger circle of the AH keeps increasing, and intersects the larger contour of
the AAH (which is not a circle) at two points at every instant.

Three-dimensional surface-plots of the AAH and AH at  $t=11.5$ and $t=13$ are
presented in Figs. \ref{fig5} and \ref{fig6}. At $t=11.5$ the Big Crunch
singularity has not yet appeared, the AAH is still a connected surface and the
AH consists of two disjoint parts -- one at $M_{AH} = 0.783$ and the other at
$M_{AH} = 3.199$. Both parts are presented in Fig. \ref{fig5} (the dotted
surfaces), and, as seen, each one has one side inside the AAH and one side
outside it (the solid surface). Each part of the AH has the shape of a sphere,
while at $t=11.5$ the AAH has the shape of a ping pong ball depressed on one
side.

At $t=13.0$ the singularity already exists at $R = 0$. Each of AH and AAH
consists only of a single surface, which surrounds the singularity. As before,
part of the AAH is outside the AH, while the other part is located inside the
AH.

Between $t = 11.5$ and $t = 13.0$ there is a period when each of AH and AAH is
split into two disjoint parts. We do not provide an illustration for this
configuration because it would be unreadable. One can imagine it as the object
from Fig. \ref{fig6} that contains a small-scale copy of itself inside. The
small object inside does not intersect with the large one.

Figures \ref{fig4} -- \ref{fig6} demonstrate that the AH and AAH do not in fact
reveal the whole truth about the future fate of the light rays. There exists a
region between the $M = 0$ axis and the AAH$-$ in Fig. \ref{fig1}, in which
future-directed rays are only formally not yet in the black hole because they
are able to proceed outwards. However, they can do so only for a short while.
They have no way to avoid intersecting the AAH and the AH in near future, and so
they are doomed to hit the Big Crunch.

\begin{figure*}
\begin{center}
  \includegraphics[scale = 0.55]{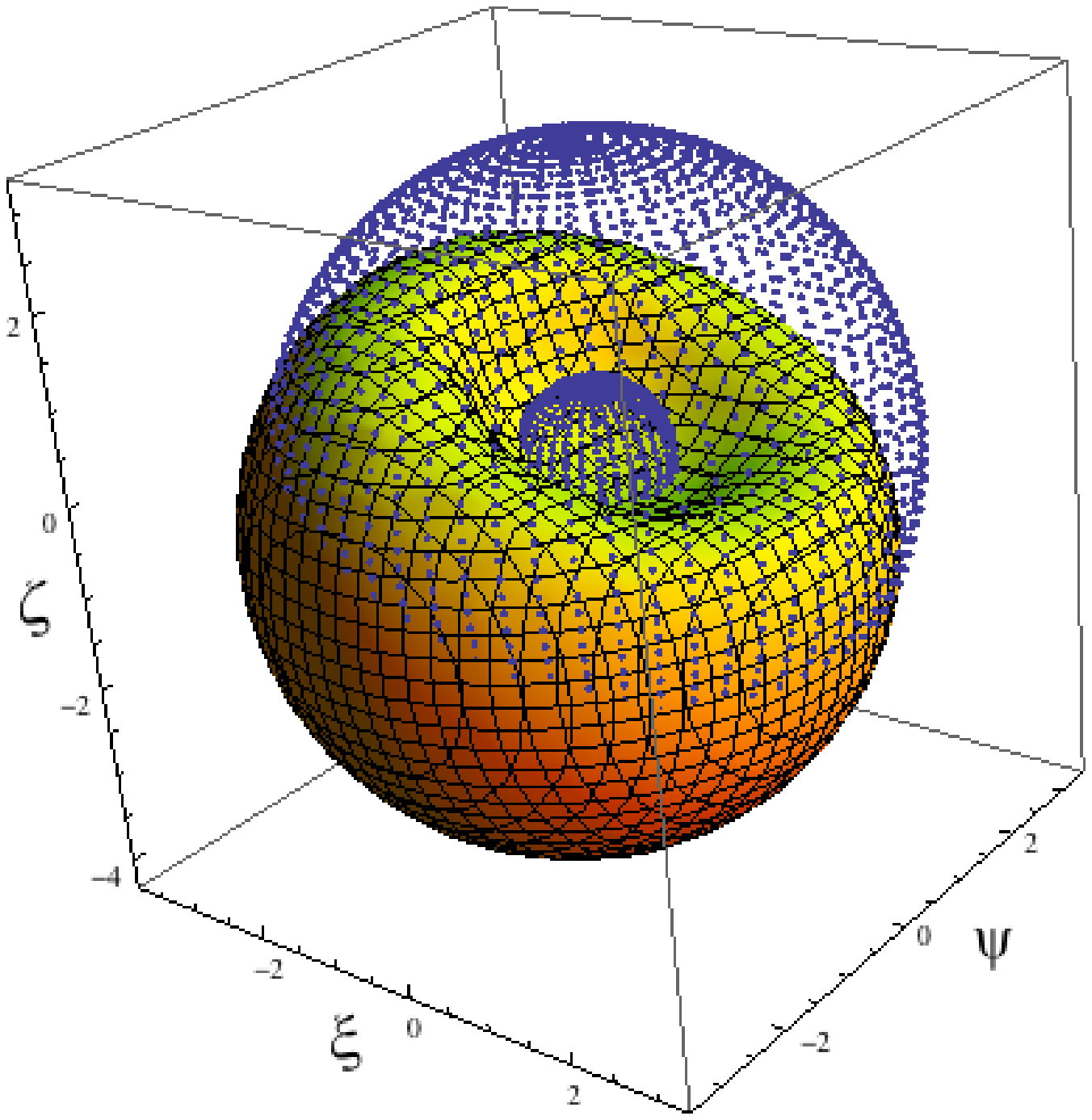}
  \includegraphics[scale = 0.65]{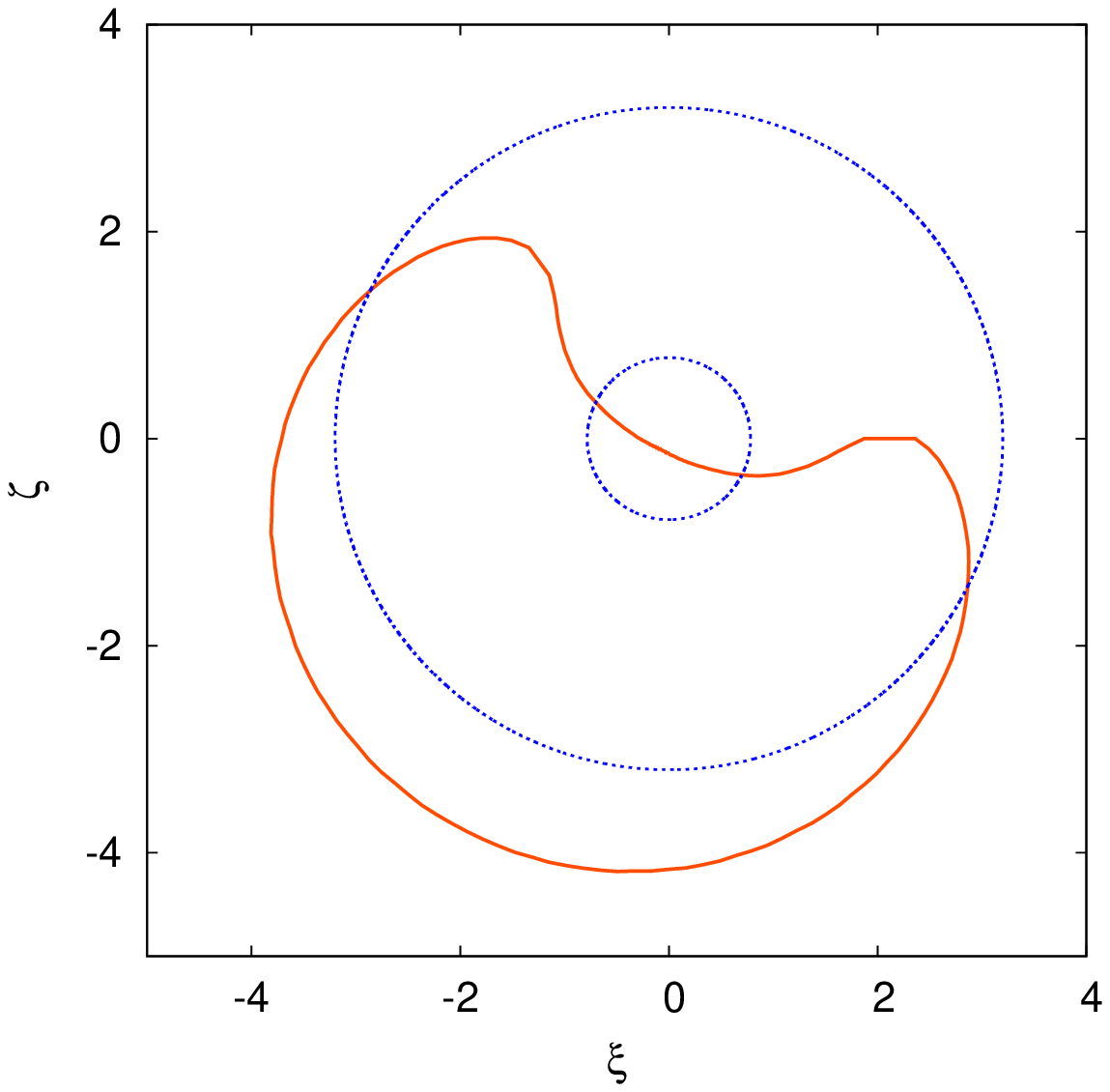}
\caption{{\em Left}: The $M$-coordinate on the AAH (solid surface) and AH
(dotted surface) represented as a function of $\vartheta$ and $\varphi$ in the
space $t = 11.5$. The value of $M(\vartheta, \varphi)$ is the distance of a
point on the surface shown from the point $(0, 0, 0)$, in the direction
specified by $(\vartheta, \varphi)$. The axes in this picture are in an abstract
Euclidean space with coordinates $(\xi, \psi, \zeta)$ used only to embed the AAH
[see transformation (\ref{6.7})]; the $(\xi, \psi)$ do not coincide with the
$(x, y)$ of Fig. \ref{fig2}. The $(M, \vartheta, \varphi)$ are spherical polar
coordinates in this space. The AH consists of two disjoint spheres -- one at
$M_{AH} = 0.783$ and the other at $M_{AH} = 3.199$. The origin $(\xi =0, \psi=0,
\zeta=0)$ is inside the smaller AAH surface. {\em Right}: Intersections of the
AAH (solid line), and the inner and outter AH (dotted lines) with the plane
$(\xi, \psi = 0, \zeta)$ (analogous to Fig. \ref{fig4}).} \label{fig5}
\end{center}
\end{figure*}

\begin{figure*}
\begin{center}
  \includegraphics[scale = 0.55]{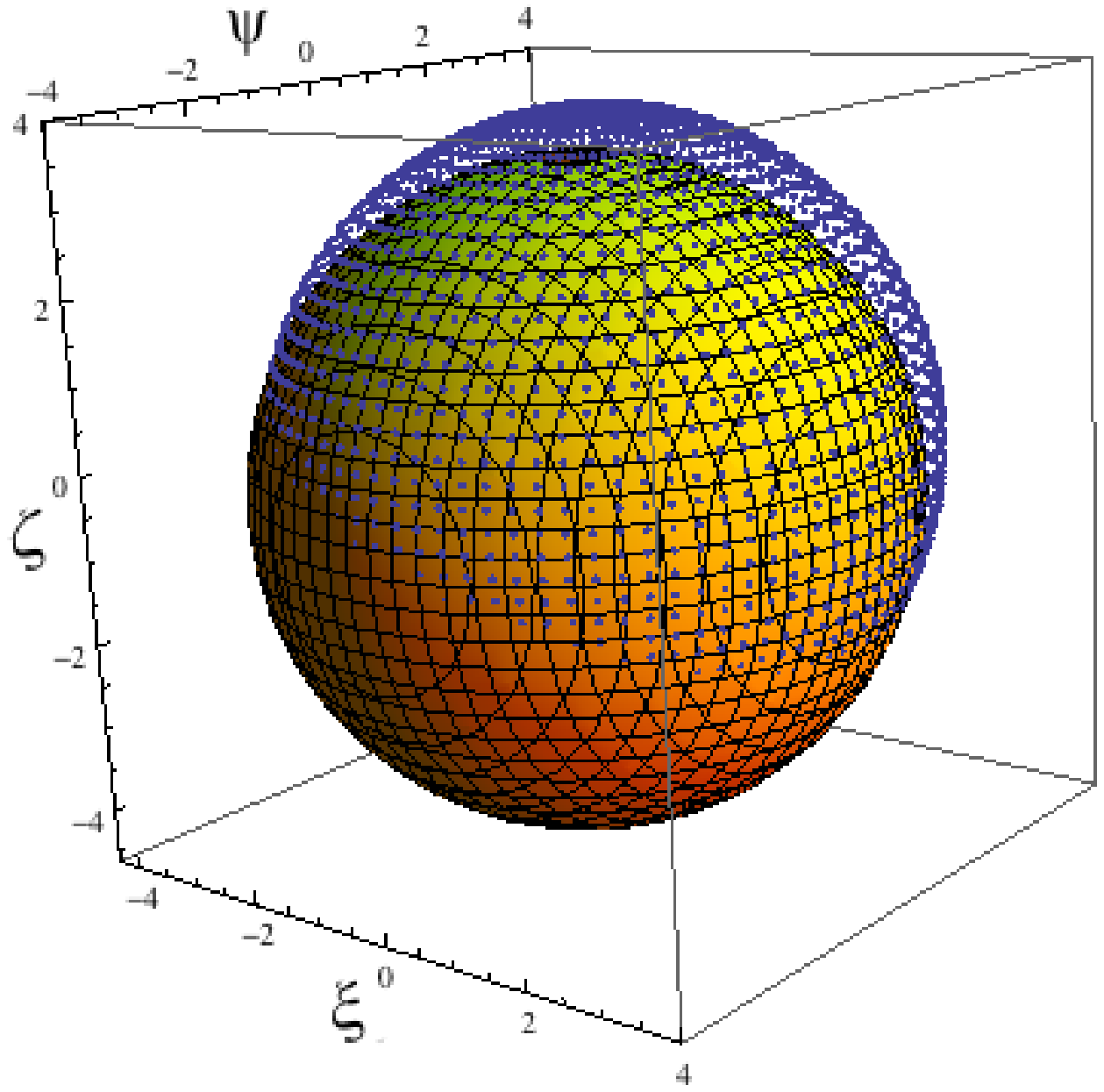}
  \includegraphics[scale = 0.65]{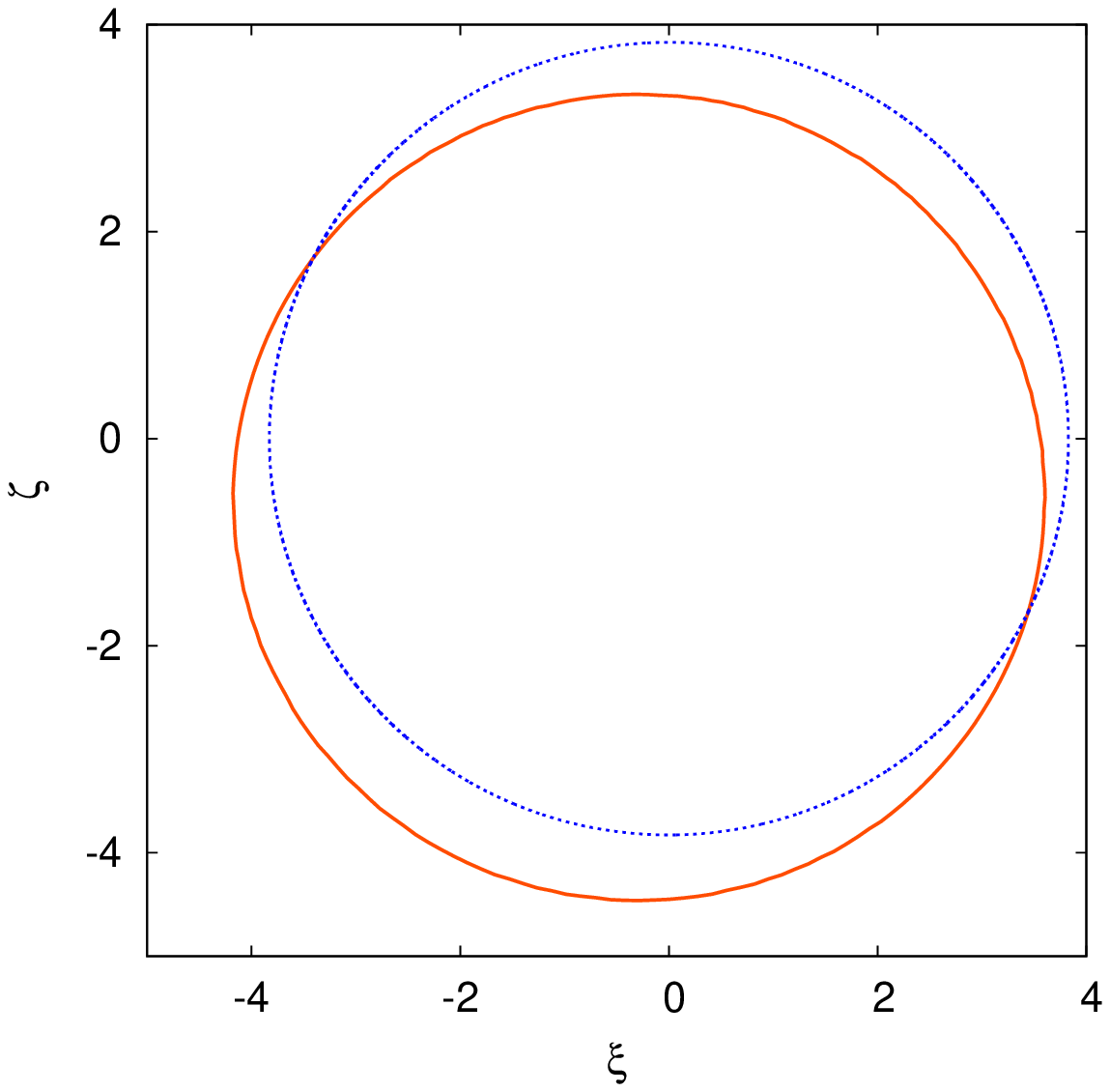}
\caption{The analogue of Fig. \ref{fig5} at the later instant $t = 13.0$, when the
singularity already exists at $R = 0$. }
\label{fig6}
\end{center}
\end{figure*}

\subsection{Timelike curves}

Equation (\ref{4.4}) for timelike trajectories is
\begin{eqnarray}\label{6.8}
&& \frac{\left(R,_z - R {\cal E},_z / {\cal E}\right)^2} {1 + 2E} \left(\dr z
t\right)^2 = \left[1 - \left(\frac{{\rm d}s}{{\rm d} t}\right)^2\right]
\nonumber \\
&& - \frac {R^2} {{\cal E}^2} \left[\left(\dr x t\right)^2 +
\left(\dr y t\right)^2\right].
 \end{eqnarray}
For a trajectory with constant $x$ and $y$ we write
\begin{equation}\label{6.9}
 \frac{\left(R,_z - R {\cal E},_z / {\cal E}\right)^2} {1 + 2E} \left(\dr z
t\right)^2 = {\cal V}^2,
 \end{equation}
where ${\cal V} = \sqrt{1 - \left(\frac{{\rm d}s}{{\rm d} t}\right)^2 }< 1$.
Then (\ref{4.8}) becomes
\begin{eqnarray}\label{6.10}
&& R,_z \left(\sqrt{1 + 2E} - \frac{\sqrt{2M/R + 2E}}{{\cal V}}\right)
\nonumber \\
&& + R \frac{\sqrt{2M/R + 2E}}{{\cal V}} {\cal E},_z / {\cal E} = 0,
\end{eqnarray}
and (\ref{4.11}) becomes
 \begin{widetext}
\begin{eqnarray}\label{6.11}
\Psi (\eta) &=& \left[\left(\frac {M,_z} M - \frac {E,_z} E\right) (1 - \cos
\eta)^{3/2} + \left(\frac 3 2 \frac {E,_z} E - \frac {M,_z} M\right) \frac {\sin
\eta (\eta - \sin \eta)} {\sqrt{1 - \cos \eta}} - \frac {(- 2E)^{3/2}} M t_{B,z}
\frac {\sin \eta} {\sqrt{1 - \cos \eta}}\right] \nonumber \\
&\times& \left[\sqrt{1 + 2E} \sqrt{1 - \cos \eta} + \frac {\sqrt{- 2E} \sin
\eta} {{\cal V} \sqrt{1 - \cos \eta}}\right] - \frac{\sqrt{- 2E}}{{\cal V}}\sin
\eta (1 - \cos \eta) \frac {{\cal E},_z} {\cal E} = 0.
\end{eqnarray}
 \end{widetext}

\begin{figure}
\begin{center}
 \includegraphics[scale = 0.7]{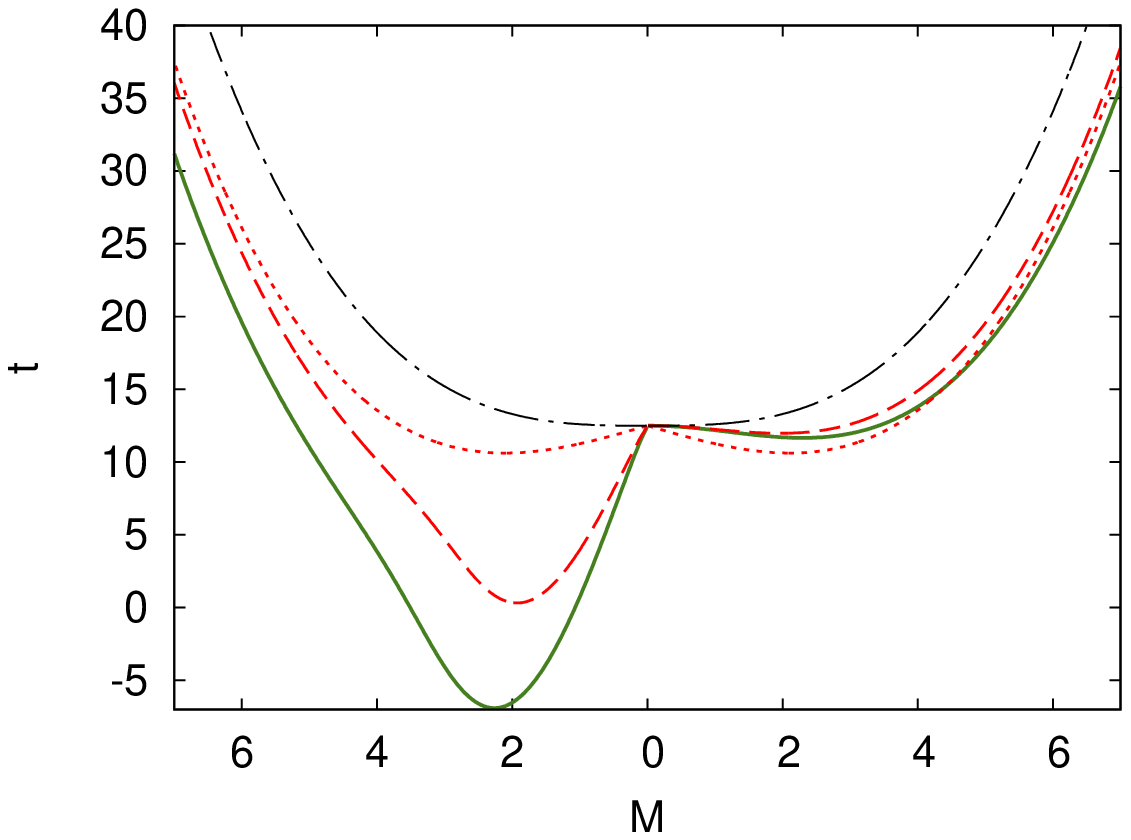}
\caption{Comparison of the AAH of a null ray (dashed lines) with the AAH of a
particle moving with velocity ${\cal V} =0.9$ (solid lines). The dotted line
represents the AH, and the dashed-dotted line represents the big crunch (BC)
singularity. As in Fig. \ref{fig1} the inner curve (closer to BC) is AAH$-$, the
outer curve is the AAH+. } \label{fig7}
\end{center}
\end{figure}

Figure \ref{fig7} is the analogue of Fig. \ref{fig1} for a particle moving with
velocity ${\cal V} =0.9$.

 \section{Which is the true horizon -- AH or AAH?}\label{truehor}

 \setcounter{equation}{0}

To get insight into this question we recall that a quasispherical Szekeres
spacetime can be matched to the Schwarzschild spacetime across an $z = b =$
constant hypersurface; this was first proved by Bonnor \cite{Bonn1976a,
Bonn1976b}. We recapitulate the basic facts about this matching by the method of
Ref. \cite{HeKr2008}.

To verify the matching, the Schwarzschild solution must first be transformed to
the Lema\^{\i}tre \cite{Lema1933} -- Novikov \cite{Novi1964} coordinates, see
Ref. \cite{PlKr2006} (section 14.12) for a derivation. In these coordinates, it
has the form
 \vspace{-3mm}
\begin{equation}\label{7.1}
{\rm d} s^2 = {\rm d} t^2 - \frac {{R,_r}^2} {1 + 2E(r)} {\rm d} r^2 - R^2(t, r)
\left({\rm d} \vartheta^2 + \sin^2 \vartheta {\rm d} \varphi^2\right),
\end{equation}
where $R(t, r)$ is determined by the equation
\begin{equation}\label{7.2}
{R,_t}^2 = 2E(r) + \frac {2m} R,
\end{equation}
$m$ being the Schwarzschild mass and $E(r)$ being an arbitrary function. In this
form, the Schwarzschild metric is the limit $M,_r = 0$ of an L--T model, and the
limit of constant $M, P, Q, S$ of a quasispherical Szekeres solution.

Further, the coordinates used on a sphere of constant $(t, r)$ in (\ref{7.1})
must be transformed to those used in (\ref{2.1}). Suppose the matching is to be
done at $r = z = b =$ constant. Then the transformation is
\begin{eqnarray}\label{7.3}
\vartheta &=& 2 \arctan \left\{\frac {\sqrt{[x - P(b)]^2 + [y - Q(b)]^2}}
{S(b)}\right\}, \nonumber \\
\varphi &=& \arctan \left[\frac {y - Q(b)} {x - P(b)}\right],
\end{eqnarray}
where $P(b), Q(b)$ and $S(b)$ are the values of the $(P, Q, S)$ from (\ref{2.1})
at $z = b$. The transformed metric (\ref{7.1}) is
\begin{equation}\label{7.4}
{\rm d} s^2 = {\rm d} t^2 - \frac {{R,_r}^2} {1 + 2E(r)} {\rm d} r^2 - \frac
{R^2(t, r)} {{{\cal E}_1}^2} \left({\rm d} x^2 + {\rm d} y^2\right),
\end{equation}
where
\begin{equation}\label{7.5}
{\cal E}_1 \df \frac {S(b)} 2 \left\{\left[\frac {x - P(b)} {S(b)}\right]^2 +
\left[\frac {y - Q(b)} {S(b)}\right]^2 + 1\right\}.
\end{equation}

Now it can be easily verified that the matching conditions between (\ref{7.4})
-- (\ref{7.5}) and (\ref{2.1}) are fulfilled at any $r = z = b =$ constant,
provided that the $E(r)$ of (\ref{7.5}) and the $E(z)$ of (\ref{2.1}) have the
same value at $r = z  = b$, and that both $R(t, b)$ are the same function of
$t$. The latter condition implies
\begin{equation}\label{7.6}
M(b)  = m,
\end{equation}
where $M$ is the function from (\ref{2.2}) and $m$ is the Schwarzschild mass
from (\ref{7.2}).

To answer the question asked in the title of this section we need to verify
whether the AH is spacelike or otherwise. For the L--T models, this analysis was
done in Ref. \cite{KrHe2004} and repeated in Ref. \cite{PlKr2006}, with the
result that the ingoing part of the AH (around $M = 0$ in Fig.
\ref{fig1})\footnote{``Ingoing'' (``outgoing'') mean, respectively, ``$R$
decreases (increases) as we proceed along the AH toward increasing $t$''.} can
be any, while the outgoing part can be spacelike and pointwise null, but never
timelike. We use the same method here, adapted to the Szekeres geometry,
assuming that (\ref{2.5}) and (\ref{2.6}) hold.

\begin{figure}
\begin{center}
 \hspace{-2.5cm}
  \includegraphics[scale = 0.6]{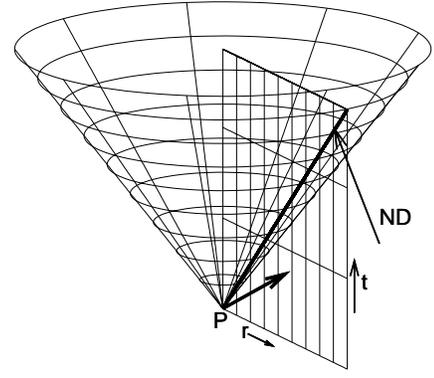}
  \includegraphics[scale = 0.8]{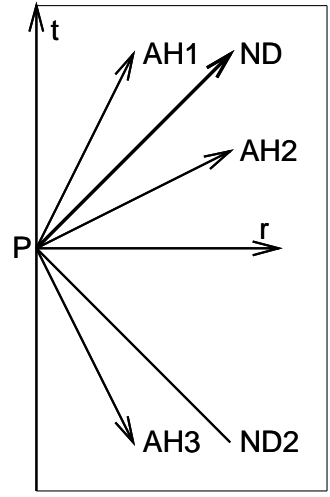}
\caption{\textit{Top:} The light cone at a point $P$ on the AH and its
intersection with the plane tangent to the $(t, r)$ surface at $P$. The
intersection determines a null direction ND -- the thick line. The arrow in the
plane marks a hypothetical direction tangent to the AH. \textit{Bottom:} The
plane from the upper panel. The figure shows also the past light cone of $P$
(ND2) and three hypothetical directions of the vector tangent to the AH at $P$.
In the position AH1, the AH would be outgoing timelike, in AH2 -- spacelike (the
position marked in the top panel), and with AH3 -- ingoing timelike. If the
direction of AH coincides with ND or ND2, then the AH is null at $P$. The
quantity $B$ defined in (\ref{7.10}) identifies the various possibilities. }
\label{fig8}
\end{center}
\end{figure}

{}From (\ref{4.3}) we find $R,_t {\rm d} t + R,_z {\rm d} z = 2 M,_z {\rm d} z$
along the AH, so
\begin{equation}\label{7.7}
\left.\dr t z\right|_{\rm AH} = \left.\frac {2M,_z - R,_z} {R,_t}\right|_{\rm
AH}.
\end{equation}
But in the collapse phase $R,_t = - \sqrt{2M/R + 2E}$, and along AH $R = 2M$, so
\begin{equation}\label{7.8}
\left.\dr t z\right|_{\rm AH} = \left.\frac {R,_z - 2M,_z} {\sqrt{1 +
2E}}\right|_{R = 2M}.
\end{equation}
The equation of the AH is independent of $(x, y)$, so a vector tangent to the AH
has only the $t$- and $r$- components. We consider an intersection of a $(t, r)$
surface with the light cone at a point of the AH, see Fig. \ref{fig8}. Along a
null geodesic that is tangent to this surface at the vertex of the light cone
(and so has $\dril x t = \dril y t = 0$) we have from (\ref{5.3})
\begin{equation}\label{7.9}
\left.\dr t z\right|_{\rm ng/AH} = \left.\frac {R,_z - 2M {\cal E},_z / {\cal
E}} {\sqrt{1 + 2E}}\right|_{\rm ng/AH}.
\end{equation}
As the lower panel of Fig. \ref{fig8} shows, the following quantity indicates
whether the AH is spacelike or otherwise
\begin{eqnarray}\label{7.10}
B &\df& \frac {(\dril t z)_{\rm AH}} {(\dril t z)_{\rm ng/AH}} = \left.\frac
{R,_z - 2M,_z} {R,_z - 2M {\cal E},_z / {\cal E}}\right|_{\rm ng/AH} \nonumber
\\
&\equiv& 1 - \left.\frac {2 \left(M,_z - M {\cal E},_z / {\cal E}\right)} {R,_z
- 2M {\cal E},_z / {\cal E}}\right|_{\rm ng/AH}.
\end{eqnarray}
Namely

\medskip

 \begin{tabular}{l|l}
THE AH IS & WHEN \\
 \hline
outgoing timelike & $B > 1$ \\
outgoing null & $B = 1$ \\
spacelike & $-1 < B < 1$ \\
ingoing null & $B = -1$ \\
ingoing timelike & $B < -1$ \\
 \end{tabular}

 \medskip

\noindent From (\ref{2.5}) and (\ref{2.6}) we see that $M,_z - M {\cal E},_z /
{\cal E} > 0$ and $R,_z - 2M {\cal E},_z / {\cal E} > 0$ are always fulfilled,
so necessarily $B < 1$, i.e. the AH can never be outgoing timelike or null; its
outgoing part is necessarily spacelike. The other three possibilities listed in
the table are allowed.

As Fig. \ref{fig1} shows, even if part of the ingoing branch of the AH is
timelike, a null curve crossing the AH outwards from the inside will be trapped
in the funnel formed by the AAH around $R = 0$. Whether it later crosses the AAH
or not, it will be forced to hit the Big Crunch within a finite segment of its
affine parameter. Where the AH is spacelike, a null line that crossed it once
cannot cross it again without being redirected toward the past. This shows that
in its outgoing part the AH acts as a black hole surface, even in that region
where the AAH is inside it. The conclusion is that the true horizon is the AH
rather than the AAH.

This conclusion is strengthened by the following consideration. If a portion of
the Szekeres manifold, of finite spatial diameter, is matched to the
Schwarzschild solution, then, from (\ref{7.6}), the AH matches to the
Schwarzschild event horizon. Thus, no signal can escape to infinity if it was
within the Szekeres AH while crossing the outer surface at $z = b$. The
intersection of the AAH with the outer surface of the Szekeres ball leaves no
trace in the Schwarzschild geometry. In particular, this happens in that part of
the Szekeres region, where the AAH is earlier than the AH (e.g. in the left half
of Fig. \ref{fig1}).

\begin{figure}[h]
\begin{center}
 \includegraphics[scale = 0.7]{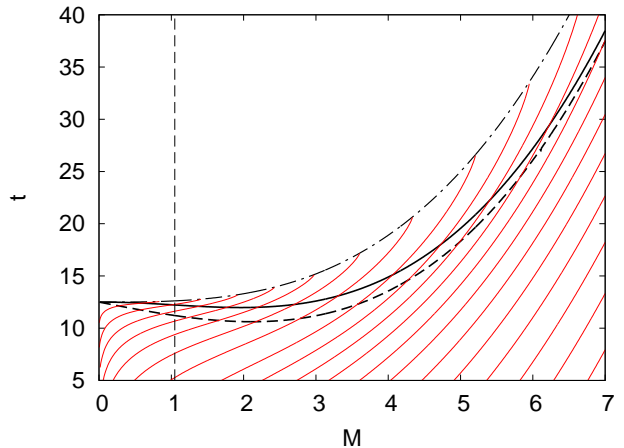}
\caption{A family of the nearly radial null rays defined by (\ref{4.5}) (thin
solid lines). The solid thick line represents the AAH$-$, the dashed thick line
represents the AH, and the dashed-dotted line represents the big crunch (BC)
singularity. To the right of the dashed vertical line ($M=1.048$) the AH is
spacelike, to the left it is ingoing timelike.} \label{fig9}
\end{center}
\end{figure}

The above considerations are illustrated in Fig. \ref{fig9}, which shows a
family of the nearly radial nongeodesic null lines (NRNL) for the model
discussed in Sec. \ref{AAHexpliA}. It is similar to Fig. \ref{fig1}, but for
clarity we only show one part of it, where the AH is below the AAH.
Calculations\footnote{At each point of intersection of the NRNL with the AH and
with the AAH we numerically calculated and compared their slopes.} showed that,
depending on the direction and the value of $M$, both the AAH and AH can be
spacelike or ingoing timelike. Figure \ref{fig9} shows the behaviour in the
direction of maximal contribution from ${\cal E}_{,z}/{\cal E}$. In this
direction the AAH$-$ is everywhere spacelike, while the AH is ingoing timelike
for $M<1.048$ (the vertical line in Fig. \ref{fig9}) and spacelike for
$M>1.048$. In other directions parts of the AAH can be timelike. The location of
the border between the ingoing timelike and spacelike parts of the AH is
direction-dependent, as follows from (\ref{7.10}). (This is so because the slope
of the null cone generator is direction-dependent, as seen from (\ref{7.9})). In
neither case can the AAH or the AH be outgoing timelike. Thus, even if in some
regions radial null rays can propagate toward increasing $R$, they cannot escape
from inside the AH and eventually they cross the AAH.

 \section{Summary}\label{conclu}

 \setcounter{equation}{0}

In order to gain a deeper insight into the quasispherical Szekeres geometries we
have investigated the spatial relation between the apparent horizon (AH) as
defined by Szekeres \cite{Szek1975b} and the absolute apparent horizon (AAH).
The concept of the AAH was first introduced by Hellaby and Krasi\'nski
\cite{HeKr2002}, but under the name of AH. In Ref. \cite{HeKr2002}, this spatial
relation was investigated at a general level, at which it was not possible to
give graphical examples. Such graphical examples are given here in the simple
subcase of the Szekeres model defined by (\ref{6.1}) -- (\ref{6.2}) and
(\ref{6.5}). The examples illustrate what was said in Ref. \cite{HeKr2002}, that
the AAH ``is a kind of oval with half inside $R = 2M$ and half outside''. The
shape of this ``kind of oval'' is shown here in Figs. \ref{fig5} and \ref{fig6}.

An observer who would fall into the region between the surfaces of AAH and AH
(top part of Fig. \ref{fig6} between the solid surface and dotted surface) for a
short while would have a chance to send a message some distance outwards (i.e on
a path with increasing $R$). This is because in a general Szekeres model the
hypersurface at which the light rays begin to converge (the AH) does not
coincide with the hypersurface at which all rays are forced to proceed toward
decreasing areal radius $R$ (the  AAH). However, the signal cannot proceed far
enough to escape the AH. Moreover, if the Szekeres spacetime is matched to the
Schwarzschild spacetime, the AH finds its prolongation in the Schwarzschild
event horizon. Consequently, it is the AH that acts as a true horizon.

Whether the concept of AAH can be usefully applied to astrophysical
considerations about galactic black holes remains to be seen. For this purpose,
the current position of the AH in space, inside the galaxy chosen for
observation, would have to be precisely determined, which seems to be a rather
remote possibility.

 \appendix

 \section{The proof that the solution of (\ref{4.11}) is unique}\label{unique}

 \setcounter{equation}{0}

We prove here that (\ref{4.11}) is fulfilled by only one value of $\eta \in
(\pi, 2\pi)$ for each set of values of $(z, x, y)$.

We begin by recalling the following:

1. Equation (\ref{4.11}) that determines $\eta(z, x, y)$ on the AAH is derived
from (\ref{4.7}), and so all quantities in it are calculated along the nearly
radial nongeodesic null line (NRNL) that obeys (\ref{4.4}) with $x$ and $y$
being constant, i.e.
\begin{equation}\label{a.1}
\frac {\left(R,_z - R {\cal E},_z / {\cal E}\right)^2} {1 + 2E} \left(\dr z
t\right)^2 = 1.
 \end{equation}

2. Note, from (\ref{2.3}), that
\begin{equation}\label{a.2}
\pdr t {\eta} = \frac M {(- 2E)^{3/2}}\ (1 - \cos \eta) > 0 \qquad {\rm for}\
\eta \in (0, 2\pi),
\end{equation}
at every fixed $(z, x, y)$, so $t(\eta)$ is a monotonic function, in this range,
i.e. $\left(\eta_i < \eta_j\right) \Longrightarrow \left(t(\eta_i) <
t(\eta_j)\right)$ for every fixed $(z, x, y)$.

Now suppose that (\ref{4.11}) has more than one solution for $\eta$ at a given
$(z, x, y)$, and call the solutions $(\eta_1, \dots, \eta_k)$, with $\eta_1 <
\dots < \eta_k$. Then (\ref{a.2}) implies that there would be $k$ instants $t_1
< \dots < t_k$, at which the given NRNL would intersect the AAH, all the $t_i, i
= 1, \dots, k$ corresponding to the same $(z, x, y)$ in (\ref{4.11}). Our
supposition thus implies that the inverse function to $t(z, x, y)$ has the
property $z(t_1, x, y) = z(t_2, x, y)$. Since this function is continuous (even
differentiable, see (\ref{a.1})), this means that for some $\overline{t} \in
(t_1, t_2)$ we have $\left.\dril z t\right|_{t = \overline{t}} = 0$. But from
(\ref{a.1}) we have
\begin{equation}\label{a.3}
\dr z t = \pm \frac {\sqrt{1 + 2E}} {R,_z - R {\cal E},_z / {\cal E}}.
 \end{equation}
This can be zero only where $E = -1/2$. This set is a neck \cite{PlKr2006} -- an
analogue of the Kruskal -- Szekeres wormhole in the Schwarzschild solution. i.e.
a special location in spacetime that may or may not exist, depending on whether
$E$ attains the value $-1/2$ anywhere.

Thus, for every NRNL obeying (\ref{a.1}) that does not traverse a neck, $\dril z
t \neq 0$ everywhere along it. This means that $z(t_1, x, y) = z(t_2, x, y)$
cannot happen, i.e. that (\ref{4.11}) has only one solution for $\eta$ at each
given $(z, x, y)$. $\square$

 \section{The limit of AAH at the center $M = 0$}\label{AAHat0}

 \setcounter{equation}{0}

For the numerical calculation we need to know the value of the function
$t(M)_{\rm AAH}$ in (\ref{4.12}) at $M = 0$. This has to be calculated exactly
because numerical programs are unreliable in calculating limits. From
(\ref{6.1}) and (\ref{6.3}) we have
\begin{equation}\label{b.1}
\lim_{M \to 0} t_B(M) = t_{B0}, \qquad \lim_{M \to 0} \frac M {(-2 E)^{3/2}} =
\frac {T_0} {2 \pi},
\end{equation}
so
\begin{equation}\label{b.2}
\lim_{M \to 0} t(M)_{\rm AAH} = t_{B0} + \frac {T_0} {2 \pi} \lim_{M \to 0}
(\eta - \sin \eta).
\end{equation}
In order to calculate $\lim_{M \to 0} \eta$ we use (\ref{4.11}). Equation
(\ref{4.15}) shows that $\sin \eta / \sqrt{1 - \cos \eta}$ is finite in the full
range $\eta \in [0, 2 \pi]$. We take $M$ as the $z$-coordinate and observe from
(\ref{2.1}) that with $(P, Q, S)$ given by (\ref{6.5}), ${\cal E},_M / {\cal E}$
will be finite at $M = 0$. Then, substituting (\ref{6.1}) and (\ref{6.3}) in
(\ref{4.11}) we get
\begin{equation}\label{b.3}
\lim_{M \to 0} \Psi(\eta) = \lim_{M \to 0} \frac {(1 - \cos \eta)^2} {3 M} = 0
\end{equation}
(because $\Psi(\eta) = 0$ all along the AAH). This is possible only when
$\lim_{M \to 0} \cos \eta = 1$, which, in the collapse phase, means
\begin{equation}\label{b.4}
\lim_{M \to 0} \eta = 2 \pi.
\end{equation}
Using this in (\ref{b.2}) we obtain
\begin{equation}\label{b.5}
\lim_{M \to 0} t(M)_{\rm AAH} = t_{B0} + T_0 = \lim_{M \to 0} t_C(M),
\end{equation}
i.e. the AAH at $M = 0$ coincides with the Big Crunch.

\bigskip

{\bf Acknowledgement:} This research was supported by the Polish Ministry of
Education and Science grant no N N202 104 838 (AK) and the European Union
Seventh Framework Programme under the Marie Curie Fellowship, grant no
PIEF-GA-2009-252950 (KB).

 \end{document}